# Magnetic field induced by elliptical instability in a rotating tidally distorded sphere

**Patrice Le Gal, Laurent Lacaze and Stéphane Le Dizès**

Institut de Recherche sur les Phénomènes Hors Equilibre, 49 rue F. Joliot-Curie, 13384, Marseille, France

E-mail : legal@irphe.univ-mrs.fr

**Abstract**. It is usually believed that the geo-dynamo of the Earth or more generally of other planets, is created by the convective fluid motions inside their molten cores. An alternative to this thermal or compositional convection can however be found in the inertial waves resonances generated by the eventual precession of these planets or by the possible tidal distorsions of their liquid cores. We will review in this paper some of our experimental works devoted to the elliptical instability and present some new results when the experimental fluid is a liquid metal. We show in particular that an imposed magnetic field is distorted by the spin-over mode generated by the elliptical instability. In our experiment, the field is weak (20 Gauss) and the Lorenz force is negligible compared to the inertial forces, therefore the magnetic field does not modify the fluid flow and the pure hydrodynamics growth rates of the instability are recovered through magnetic measurements.

## 1. The elliptic instability of rotating flows

1.1 *Experiments in deformed cylinders.* Motivated by the analyses of elliptical instabilities of shear flows or vortices [1], we performed an exhaustive theoretical and experimental study of the unstable modes of a deformable fluid cylinder of radius r, in rotation around its axis at a rate $\Omega$. These modes whose existence is related to resonances of inertial waves have been fully characterized by theoretical analyses (frequencies, wave numbers, shapes) and compared to experimental results [2]. Moreover, we highlighted that at low Ekman numbers ($E=\nu/\Omega r^2$), intermittent cycles between instability growth periods, turbulent breakdown and relaminarisation windows can occur. These intermittent cycles have already been observed by Malkus [3] who related them to the "resonant collapses" of inertial waves when excited in rotating containers [4]. Figure 1 illustrates one of this intermittency where it is even possible that the phase of the unstable mode reverses between two cycles. Laser Doppler anemometry measurements of the axial velocity in such a cylinder confirms these intermittent cycles. Figure 2 presents a time series of the axial velocity at some location in the cylinder. A clear chaotic intermittent behavior is observed. However, we have not yet been able to determine the precise scenario of transition to chaos in this flow. One of the difficulty comes from the poor resolution of our measurements which is mainly due to the bad optical property of the deformable rotating cylinder wall. We will see later that there exists however a possibility to probe the velocity field by measuring the distorsions of an imposed magnetic field.

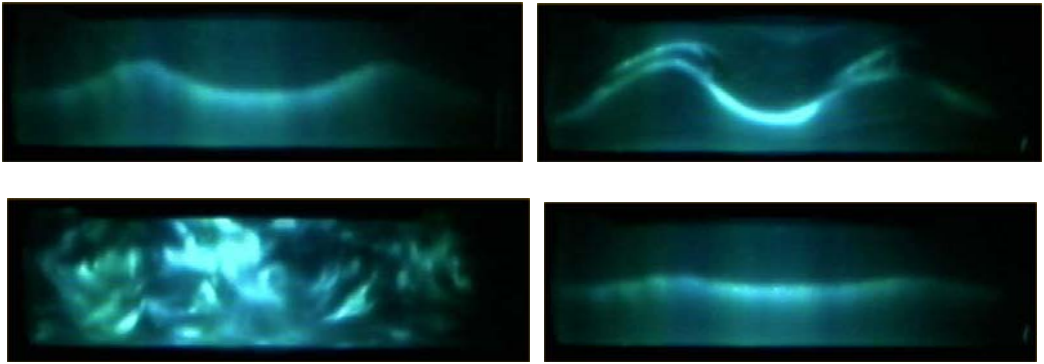

**Figure 1**: Four successive snapshots of the intermittent cycles of the elliptical instability in a cylinder.

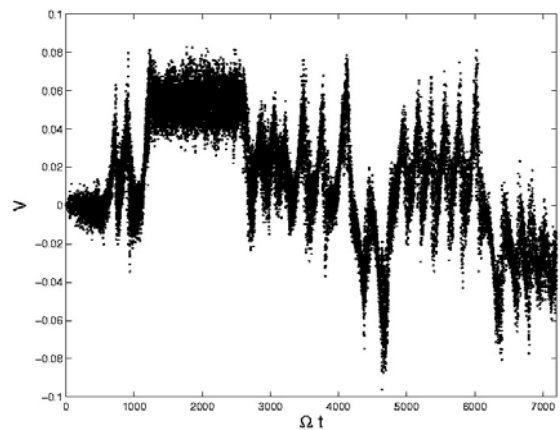

**Figure 2**: Time series of the axial velocity (m/s) inside the cylinder. The intermittent behavior is clearly visible; some "reversals" of the phase of the unstable mode (negative velocities) are particularly visible.

1.2 *Experiments in deformed spheres.* In a second series of experiments, a slightly deformed sphere is set into rotation. The experimental device was already described by Lacaze et al. [6]. A ping pong ball ball is inserted in a cylindrical block of liquid silicone that is cured at a temperature of 50° Celsius with the ping-pong ball inside. Finally, the ball is dissolved by a solution of ethyl acetate and a hollow sphere molded in a transparent and deformable cylinder is obtained. The radius of the sphere is 21.75mm. The silicone cylinder is mounted on the vertical shaft of the device already used in the cylindrical case and is compressed between two vertical rollers. Note that these rollers are always in place; the device does not offer the possibility to move them in or out after rotation is started. The distance separating these rollers gives directly the elliptical deformation $\varepsilon$ of the deformable sphere.

The spherical geometry is obviously interesting in a geophysical context. Indeed, the liquid molten iron cores of planets can be deformed in an elliptic shape under the effect of tides generated by the gravitational field of their suns or moons. This deformation which is for instance about forty centimeters in the case of the Earth and of several hundreds of meters for the Jovian satellite Io, is then suitable to trigger inertial waves resonances in rotating planetary cores [5, 7]. The resulting fluid motions inside these liquid metal cores could then may have some relevance for the generation and the dynamics of planetary magnetic fields.

As observed in figure 3-a), an unstable mode called the "spin-over", which forces the fluid to turn around an axis perpendicular to the rotation entrainment axis, gives an "S" shape to the fluid axis of rotation. A theoretical analysis of the instability in a spherical geometry has also been carried out. It confirms the appearance of the spin over mode whose growth rates (see figure 3-b) and non linear saturation can be advantageously compared to the measurements [5].

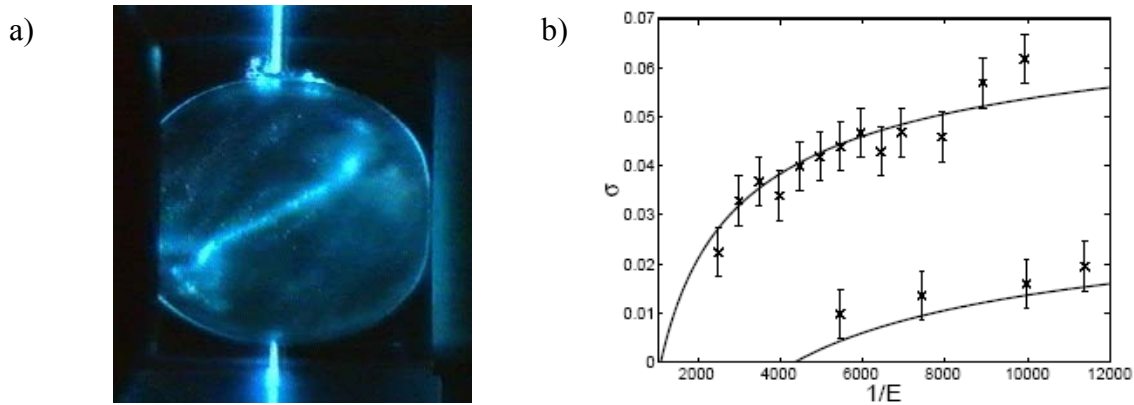

**Figure 3** a) Spin-over mode generated by the elliptic instability in a deformed rotating sphere. b) comparison between theory and experimental measurement of the growth rates of the spin-over mode for two cases of ellipticity ($\varepsilon = 0.08$ and $0.16$).

## 2. Induction of a magnetic field

We then studied the response of the flow to an imposed magnetic field. A field $B_0$ of 20 Gauss was created by the use of two Helmholtz coils (see figure 4). $B_0$ is parallel to the axis of rotation of the sphere. The sphere is filled with a Gallium-Indium-Tin eutectic (Galinstan) which is liquid at ambient temperature contrary to pure Gallium. The compression of the sphere by the rollers gives an ellipticity of $\varepsilon = 0.092$. The entrainment axis of the sphere being parallel to the field $B_0$, it is not distorted by this solid rotation. But as the spin-over mode is a solid rotation perpendicular to $B_0$, a torsion of the magnetic field lines by the elliptical instability is produced and an induced horizontal field $b_H$ is generated. The schematic representation of figure 5 shows this effect which is in fact the clasical $\Omega$ effet of the MHD literature. Note that the magnetic Reynolds number of our flow is of the order of $10^{-2}$.

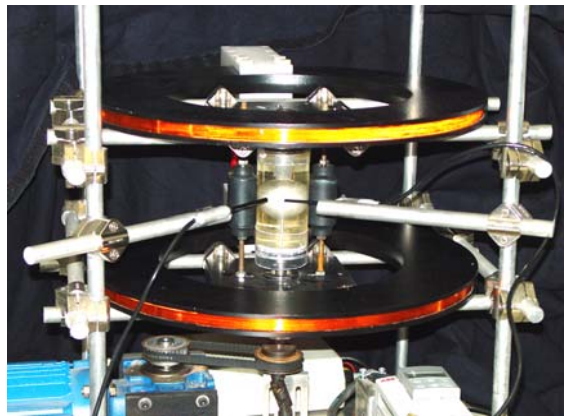

**Figure 4**: Experimental set up with the pair of Helmholtz coils surrounding the rotating sphere.

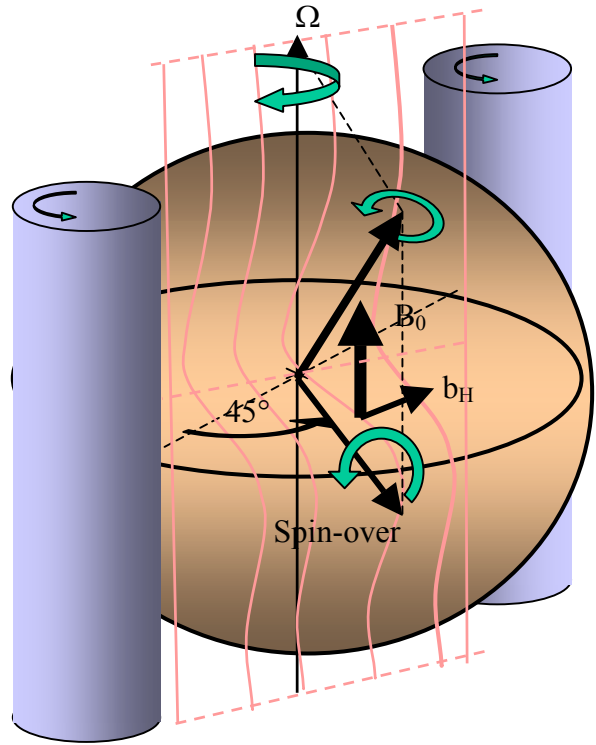

**Figure 5**: Schematic representation of the distorted magnetic field lines associated with the Ω effect generated by the spin-over mode of the elliptical instability in the rotating deformed sphere.

Using a Gaussmeter with Hall effect probes, we measured in the equatorial plane, 5 mm outside of the sphere, the growth of the magnetic field $b_H$ induced perpendicularly to the imposed field. As it can be seen in figure 6, after some oscillations coming from perturbations of the magnetic field at the onset of rotation, a clear exponential increase is measured. Then, after an overshoot, a non linear saturation of the field is observed. No more complex dynamics has yet been recorded.

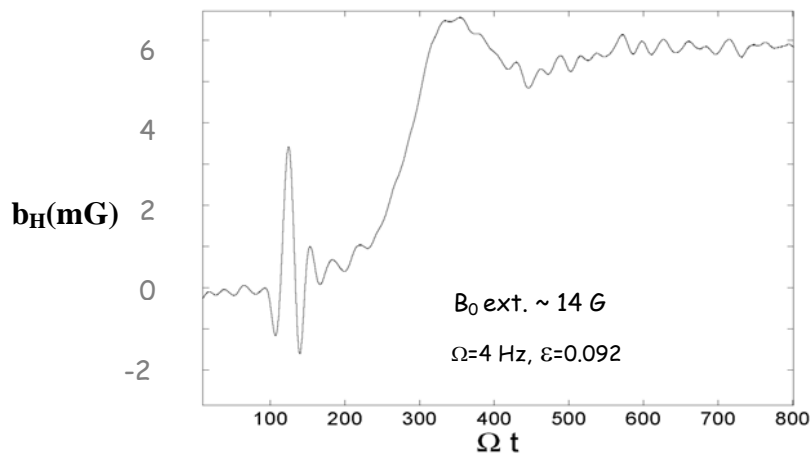

**Figure 6**: Temporal exponential growth of the induced magnetic field (mGauss). The vertical imposed magnetic field had a value of 14 Gauss at the external position of the probe.

These measurements were then repeated for various Reynolds number and semi-log plots of these transients permit the determination of the growth rates of the induced magnetic field $b_H$. A comparison of these growth rates σ with their theoretical predictions that take into account only the hydrodynamics effects is excellent as it can be seen in figure 7.

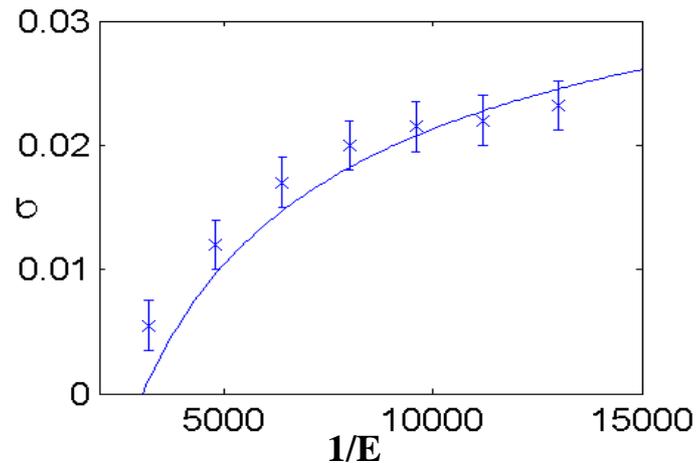

**Figure 7**: Comparison of the measured growth rates σ ( ε= 0.092) of the induced magnetic field with the theoretical predictions for ε= 0.095. The Lorenz force is not taken into account in the calculation.

## 3. Conclusion

We have proved in this study that the elliptical or tidal instability, and in particular the spin-over mode which is the most unstable mode in the slightly deformed sphere, can induce a magnetic field when the working fluid is a liquid metal and when an external field is imposed. As shown in [6] this situation is reminiscent of the Jovian satellite Io whose magnetic field origin is still unknown. Indeed, the huge tides created by Jupiter proximity can excite a resonant process similar to the one we have experimented here. The magnetic field of Jupiter could then be distorted by the core motions of Io. As calculated by Kerswell and Malkus [6], an intrisic dynamo is not needed to interpret the magnetic anomaly measured by Gallileo [8]. Nevertheless, as shown by Tilgner [9] in a numerical study of a precession flow in a sphere, it is worth noting that a resonant interaction between inertial waves, as the one we have explored here, may indeed be an alternative mechanism to generate geo-dynamos.

We would like to thank M.P. Chauve for his help during the magnetic measurement set up.

**Reference**
[1] Kerswell R R 2002 Elliptical instability *Annual Review of Fluid Mechanics* **34** pp 83-113
[2] Eloy C, Le Gal P and Le Dizès S 2003 Elliptic and triangular instabilities in rotating cylinders *J Fluid Mech* **476** pp 357-388
[3] Malkus W V R 1989 An experimental study of global instabilities due to tidal (elliptical) distortion of a rotating elastic cylinder *Geophys Astrophys Fluid Dynamics* **48** 123–134
[4] McEwan A D 1970 Inertial oscillations in a rotating fluid cylinder *J Fluid Mech* **40** pp 603–640
[5] Kerswell R R , Malkus W V R 1998 Tidal instability as the source for Io's magnetic signature *Geophys Res Lett* **25** pp 603–606
[6] Lacaze L, Le Gal P and Le Dizès S 2004 Elliptical instability in a rotating spheroid *J Fluid Mech* **505** pp 1-22
[7] Aldridge K D  Seyed-Mahmoud B  Henderson G & Van Wijngaarden W 1997 Elliptical instability of the earth's fluid core *Phys Earth Planet Int* **103** pp 365–374
[8] Kivelson M G, Khurana K K , Walker R J , Russel C T, Linker J A, Southwood D I and Polanskey C 1996 A magnetic Signature at Io: Initial Report from the Galileo Magnetometer *Science* **273** pp 337-340
[9] Tilgner A 2005  Precession driven dynamos *Phys Fluids* **17** 034104